\documentclass[12pt]{article}
\usepackage{epsfig}
\usepackage{float}
\usepackage{amssymb,amsmath}
\newcommand{\be}{\begin{equation}}
\newcommand{\ee}{\end{equation}}
\newcommand{\bea}{\begin{eqnarray}}
\newcommand{\eea}{\end{eqnarray}}
\begin{document}
\begin{center}
{\bf DIFFERENT SCHEMES OF NEUTRINO OSCILLATIONS IN M\"OSSBAUER
NEUTRINO EXPERIMENT}
\end{center}
\begin{center}
S. M. Bilenky
\end{center}

\begin{center}
{\em  Joint Institute for Nuclear Research, Dubna, R-141980,
Russia\\}
\end{center}
\begin{center}
F. von  Feilitzsch and W. Potzel
\end{center}
\begin{center}
{\em Physik-Department E15, Technische Universit\"at M\"unchen,
D-85748 Garching, Germany}
\end{center}
\begin{abstract}
We comment on the paper "On application of the time-energy uncertainty
relation to M\"ossbauer neutrino experiments" (see arXiv: 0803.1424) in
which our paper "Time-energy uncertainty relations for neutrino
oscillation and M\"ossbauer neutrino experiment" (see arXiv: 0803.0527)
has been criticized. We argue that this critique is a result of
misinterpretation: The authors of (arXiv: 0803.1424) do not take
into account (or do not accept) the fact that at present there exist
different schemes of neutrino oscillations which can not be
distinguished in usual neutrino oscillation experiments. We stress
that a recently proposed M\"ossbauer neutrino experiment provides the
unique possibility to discriminate basically different approaches to
oscillations of flavor neutrinos.
\end{abstract}

Recently Akhmedov et al. published a paper in the arXiv "On application of the
time-energy uncertainty relation to M\"ossbauer neutrino experiments"
\cite{Akhmedov1} in which they criticize our arXiv paper "Time-energy
uncertainty relations for neutrino oscillation and M\"ossbauer
neutrino experiment"\cite{BilFeilPotz08}. In this arXiv paper
we showed that the time-energy uncertainty relation does not allow
neutrino oscillations in the M\"ossbauer neutrino experiment proposed in
\cite{Raghavan} and \cite{Potzel} for the case of evolution of the neutrino state in time. We  comment here on the paper by Akhmedov et
al. \cite{Akhmedov1}.
\begin{enumerate}
  \item
The authors of \cite{Akhmedov1} write "...the conclusions of
\cite{BilFeilPotz08} are in conflict with the results of our recent detailed
quantum field theoretical calculation ..." \cite{Akhmedov2}.  This is
not the case. We are considering different schemes of neutrino
oscillations. The standard theory of neutrino oscillations in vacuum
is based on the notion of mixed flavor neutrino states $|
\nu_{e}\rangle,| \nu_{\mu}\rangle, | \nu_{\tau}\rangle $. There are
two different approaches to the evolution of the states of the
produced flavor neutrinos: evolution in time or evolution in space
and time. Both approaches give identical results for flavor
neutrino transition probabilities in usual neutrino oscillation
experiments in which there are no constraints on energies and momenta
of  neutrinos with different masses. In our paper
\cite{BilFeilPotz08} we derive the time-energy uncertainty relation for
neutrino oscillations and show that this relation does not allow
oscillations in the M\"ossbauer neutrino experiment in which
practically monochromatic neutrinos are produced.  For the
derivation of the time-energy uncertainty relation we used the general
Mandelstam-Tamm method \cite{TammMand45} which implies evolution in
time (see below). However,  in paper \cite{BilFeilPotz07} we
considered both approaches to the evolution of the flavor neutrino
states. We have shown that in the case of evolution in space and
time, oscillations of  M\"ossbauer neutrinos are perfectly possible
(due to the difference of momenta of neutrinos with the same energy and
different masses).

In paper \cite{Akhmedov2}, oscillations of M\"ossbauer neutrinos
are considered as an effect of  mixing and propagation of virtual
neutrinos between source and detector. This approach allowed
Akhmedov et al. not only to take into account the effect of neutrino
propagation but also to investigate in detail effects of neutrino
production and detection. It was shown in \cite{Akhmedov2} that
oscillations of M\"ossbauer neutrinos are possible and that the neutrino
survival probability coincides with the probability which can be
obtained in the space-time scenario.
\item It was shown in \cite{TammMand45}  that for any Heisenberg operator $O(t)$
and any Heisenberg state $| \Psi\rangle_{H} $ the following
inequality is valid
\begin{equation}\label{timeenergy1}
    \Delta E~\Delta O(t) \geq \frac{1}{2}
   ~ |\frac{d }{d t}\langle\Psi_{H}| O(t)|\Psi_{H}\rangle|
\end{equation}
In order to derive the time-energy uncertainty
relation for neutrino oscillations from this inequality we have
chosen in \cite{BilFeilPotz08}
\begin{equation}\label{proj}
O(t)=P_{l}(t), \quad (l=e,\mu,\tau)
\end{equation}
where $P_{l}(t)$ is the operator of the projection on the flavor
neutrino state $|\nu_{l}\rangle$. If the initial neutrino state is
$|\nu_{l}\rangle$ we have
\begin{equation}\label{probabil}
\langle\Psi_{H}| P_{l}(t)|\Psi_{H}\rangle=P_{\nu_{l}\to \nu_{l}}(t),
\end{equation}
where $P_{\nu_{l}\to \nu_{l}}(t)$ is the probability of $\nu_{l}$ to
survive. With this choice the inequality (\ref{timeenergy1}) takes
the form
\begin{equation}\label{timeenergy2}
\Delta E \geq \frac{1}{2}~\frac{|\frac{d }{d t}P_{\nu_{l}\to
\nu_{l}}(t) |}{\sqrt{P_{\nu_{l}\to \nu_{l}}(t)-P^{2}_{\nu_{l}\to
\nu_{l}}(t)}}~.
\end{equation}
Akhmedov et al. \cite{Akhmedov1}, using the same procedure, came to a
relation which differs from (\ref{timeenergy2}) by the change
\begin{equation}\label{change}
P_{\nu_{l}\to \nu_{l}}(t)\to P_{\nu_{l}\to \nu_{l}}(x,t).
\end{equation}
They claim  "An important point is that $P_{\nu_{l}\to \nu_{l}}$
depends in general on both $t$ and $x$". On the basis of the
modified relation (\ref{timeenergy2}), in which the change (\ref{change}) was taken into account, Akhmedov et al. conclude that the
time-energy uncertainty relation "does not preclude oscillations of
M\"ossbauer neutrinos".

In the framework  of the quantum field theory  it is impossible to
come to the conclusion that the survival probability given by
(\ref{probabil}) depends on $t$ and $x$. In fact,  for the operator
$P_{l}(t)$ in the Heisenberg representation we have
\begin{equation}\label{Ho}
P_{l}(t)=e^{iHt}~P_{l}~e^{-iHt},
\end{equation}
where $P_{l}=|\nu_{l}\rangle\langle\nu_{l}|$ is the projection
operator in the Schr\"odinger representation. This operator can not
depend on $x$ (flavor neutrino states are the same in all points of
the space). The constant Heisenberg state $|\Psi_{H}\rangle$ also
can not depend on $x$. Thus, the survival probability given by
(\ref{probabil}) can depend only on $t$. The Mandelstam-Tamm method
can not be applied to the survival probability which depends on $t$
and $x$.

\item The survival probability is determined in \cite{Akhmedov1} by
the relation
\begin{equation}\label{Ahrel}
P_{\nu_{l}\to \nu_{l}}(x,t)=|\langle\nu_{l}|\Psi(x,t)\rangle|^{2}
\end{equation}
"with $\Psi(x,t)$ being the neutrino wave function" (presumably
neutrino state vector).  We would like to comment on the statement (which has been
taken as granted in \cite{Akhmedov1}) that the neutrino state depends on
$x$ and $t$. In field theory, such a statement can not be
justified. Flavor neutrino states are states with definite momenta.
Such states can not depend on $x$.\footnote{See, for example,
ref. \cite{BogShirkov} chapter 2, p.89: "... in the quantum field theory
the state does not depend explicitly on the coordinate $x$". Let us
also notice  that the operator $e^{-iPx}$ (here $x=(t,\vec{x})$, and
$P$ is  the total momentum) is not an operator of the evolution of the
states. This operator determines the $x$-dependence of field operators:
$\psi(x)=e^{iPx}\psi(0)e^{-iPx}$, $\psi(x)$ being  {\em a field
operator.}}

\item In order
to obtain the time-energy uncertainty relation we integrate the
relation (\ref{timeenergy2}) over $t$. It was argued by Akhmedov et
al. (on the basis of the suggestion that the survival probability
depends on $x$ and $t$) that in the integration (\ref{timeenergy2})
over $t$ we used the relation
\begin{equation}\label{xt}
    x\simeq t.
\end{equation}
From the previous discussion it is obvious, however, that we do not
 use the relation (\ref{xt}) in order to perform the integration:
the survival probability in (\ref{timeenergy2}) depends only on $t$ (as
it was stressed several times in \cite{BilFeilPotz08}).

The time-energy uncertainty relation means that there exists a
correlation between the uncertainty in energy $\Delta E$ and the
"characteristic time" $ t$ of a process
\begin{equation}\label{t-e1}
\Delta E~ t \gtrsim 1.
\end{equation}
The integration of (\ref{timeenergy2}) over the time  gives us the
possibility to connect $t$ with the survival probability (see Eq. (35) in
\cite{BilFeilPotz08}). It is natural to determine the characteristic
time as a time interval during which the survival probability is
significantly changed (say, reaches the first minimum). In order to
find the characteristic time we use the standard expressions for the
survival probabilities, which can be obtained from different
approaches to neutrino oscillations assuming the relation
(\ref{xt}). As well known, the standard transition
probabilities perfectly describe the existing experimental neutrino
oscillation data.

It was proposed in \cite{Akhmedov1} to use directly the relation
(\ref{timeenergy2}) (with the change $P_{\nu_{l}\to \nu_{l}}(t)\to
P_{\nu_{l}\to \nu_{l}}(x,t)$). We would like to notice the following:
\begin{itemize}
  \item
If we use (\ref{timeenergy2}) and the standard expression for the
survival probability  for $\nu_{\mu}\to\nu_{\mu} $ transitions
with $\sin^{2}2\theta_{23}\simeq 1$ we obviously come to the relation
(40) of \cite{BilFeilPotz08}). However, in cases of non-maximal
mixing we need to use the integrated relations.
\item
The relation
\begin{equation}\label{At-e}
    \Delta E\geqslant |E_{1}-E_{2}|,
\end{equation}
which was obtained in \cite{Akhmedov1}, does not have the form of
a time-energy uncertainty relation: there is no characteristic
time in it. It is obvious that this relation is satisfied for
neutrino oscillations for the case of the evolution of flavor
neutrino states in space and time (in fact, (\ref{At-e}) was
obtained in \cite{Akhmedov1} for this case). However, the relation
(\ref{At-e}) is not satisfied for M\"ossbauer neutrinos in the case
of evolution in time (the possibility which was not considered in
\cite{Akhmedov1}).
\end{itemize}

\item We would like to comment the relation (\ref{xt}).
 Of course, this relation is not an exact
one: neutrino masses are different from zero and neutrino wave
packets have final sizes. However, for ultra-relativistic neutrinos,
corrections to (\ref{xt}) are extremely small. (Neutrino masses
are small and the sizes of wave packets of neutrinos which are produced
and detected in elementary particle (not M\"ossbauer) processes presumably have
microscopic size). Let us stress that the validity of relation
(\ref{xt}) for neutrino oscillations was confirmed by the K2K \cite{K2K} and
MINOS \cite{Minos} accelerator experiments in which the times of neutrino production
and detection were measured.

Akhmedov et al. claim that the "neutrino arrival time is not well defined
for M\"ossbauer neutrinos" and relation (\ref{xt}) can not be used.
Their argument is based on the assumption that the lengths of the wave
packets in the case of  M\"ossbauer neutrinos have macroscopic size,
much larger than the source-detector distance.

We would like to notice that Akhmedov et al. themselves use the
relation (\ref{xt}) for M\"ossbauer neutrinos. In fact, let us
compare the papers \cite{Akhmedov1} and \cite{Akhmedov2}. In
\cite{Akhmedov1}, for the oscillation phase difference the following
expression was used
\begin{equation}\label{oscphase1}
 2\phi_{jk}=(E_{j}-E_{k})t -(p_{j}-p_{k})x.
\end{equation}
After simple algebra from (\ref{oscphase1}) we find
\begin{equation}\label{oscphase2}
2\phi_{jk}\simeq (E_{j}-E_{k})(t-x)+\frac{\Delta m^2_{jk}}{2E}x.
\end{equation}
On the other hand, in \cite{Akhmedov2} it was shown that neutrino
oscillations in the M\"ossbauer neutrino experiment are described by
the standard oscillation formula with  the phase difference
\begin{equation}\label{oscphase}
 2\phi_{jk}=\frac{\Delta m^2_{jk}}{2E}x.
\end{equation}
It is obvious that equations (\ref{oscphase2}) and (\ref{oscphase})
 are compatible only if the relation (\ref{xt}) holds.

\item In
\cite{BilFeilPotz08} we wanted to address the physical
question which is still open today: are neutrino oscillations a non-stationary phenomenon? If this is the case,
the time-energy uncertainty relation (\ref{t-e1}) has to be valid for neutrino oscillations.
It is natural to expect (and we showed this in \cite{BilFeilPotz08})
that the characteristic time for the case of neutrino
oscillations is determined by  the oscillation time. In the case of M\"ossbauer neutrinos, the left-hand side of (\ref{t-e1}) is
about $4\cdot 10^{-4}$. Even if the "arrival time for M\"ossbauer
neutrinos is not well defined" \cite{Akhmedov1} it is impossible to
"overcome" such a small factor and fulfill the time-energy
uncertainty relation.

From our point of view, our discussion with the authors of the paper
\cite{Akhmedov1} is an illustration of the fact that the final theory of
neutrino oscillations still does not exist. We would like to stress
again that the different approaches to neutrino oscillations can not be
distinguished in usual neutrino oscillation experiments. The
condition of a very small energy uncertainty of neutrinos, which is
required for the M\"ossbauer resonance,  makes the proposed
M\"ossbauer neutrino experiment a unique one. There are no doubts
that if the M\"ossbauer neutrino experiment could be performed it
would have an enormous impact on our understanding of the basic physics
of neutrino oscillations.
\end{enumerate}


\begin{thebibliography}{99}
\bibitem{Akhmedov1}
E.Kh. Akhmedov, J. Kopp, and M. Lindner, arXiv: 0803.1424.

\bibitem{BilFeilPotz08} S.M. Bilenky, F. von Feilitzsch, and W.
Potzel, arXiv: 0803.0527v1.

\bibitem{Raghavan} R.S. Raghavan, arXiv: hep-ph/0601079.


\bibitem{Potzel}W. Potzel, Phys. Scripta {\bf T127} (2006) 85.


\bibitem{Akhmedov2}
E.Kh. Akhmedov, J. Kopp, and M. Lindner, arXiv: 0802.2513.

\bibitem{TammMand45}L. Mandelstam and I.E. Tamm, J. Phys.(USSR)
\textbf{9} (1945) 249.



\bibitem{BilFeilPotz07} S.M. Bilenky, F. von Feilitzsch, and W.
Potzel, J. Phys. \textbf{G34} (2007) 987.




\bibitem{BogShirkov} N.N. Bogoliubov and D.V. Shirkov,
Introduction to the theory of quantized fields, John Wiley and Sons,
1979.


\bibitem{K2K} K2K Collaboration, M.H. Alm {\em et al.},
Phys. Rev. Lett. \textbf{90} (2003) 041801.


\bibitem{Minos} MINOS Collaboration, D. G. Michael {\it et al.},
arXiv: hep-ex/0607088.


\end{thebibliography}
\end{document}